\documentstyle[twocolumn,aps,epsbox]{revtex}       

\makeatletter
\def\gsim{\mathrel{\mathpalette\@versim>}}
\def\lsim{\mathrel{\mathpalette\@versim<}}
\def\@versim#1#2{\lower3\p@\vbox{\baselineskip\z@skip\lineskip0\p@
    \ialign{$\m@th#1\hfil##\hfil$\crcr#2\crcr\sim\crcr}}}
\makeatother
\def\mathrm#1{{\rm #1}}

\def\mib#1{{\mbox{\boldmath $#1$}}}

\begin{document}

\noindent
{\bf Ikeda and Ohashi reply:}
In our previous letter \cite{rf:Ikeda}, we have proposed an unconventional
spin density wave state as a possible mechanism of the micromagnetism
in URu$_2$Si$_2$.
As an example, we have studied the $d$-wave SDW ($d$SDW).
This novel SDW can explain various experimental results.
Kiselev and Bouis (KB) have pointed out that the ferromagnetic (FM) state
should be considered in the phase diagram (Fig.1 in \cite{rf:Ikeda})
and the $d$SDW cannot be realized for the most physically reasonable limits.

In \cite{rf:Ikeda}, we analized the simplest model for the $d$SDW
(Eq.(1) in \cite{rf:Ikeda}) within the mean field theory.
It was implicitly assumed that the antiferromagnetic state in URu$_2$Si$_2$
originates from the nesting in the heavy fermion state \cite{rf:Rozing}.
Then, among the possible orderings, we examined only states with
the nesting vector $\mib{Q}$ ($Q$-group).
The states in this group are expected to always compete with one another
irrespective of the detail of models whenever the nesting works relevently.
On the other hand, since the exchange term $J$ favours the FM state,
Fig.1 in \cite{rf:Ikeda} is modified as pointed out by KB
(See Fig.1.) when possibility of the FM state is included.
We, however, note that this FM instability mainly comes from the peculiarity
of our simple model besides the presence of $J$, i.e., the divergence
of the density of states (DOS) at $E=0$.
Actually, no precursor of the FM instability has been observed
experimentally in pure URu$_2$Si$_2$ \cite{rf:Maple,rf:Torikachvili}.
In this regard, our model in \cite{rf:Ikeda} is too simple to
correctly describe this feature in real URu$_2$Si$_2$, although
it is enough to grasp the essence of the $d$SDW.
In a more realistic model \cite{rf:Fawcett}, the FM instability is expected
to be less dominant compared with the simple one.

Next, we discuss the stable region of the $d$SDW within Eq.(1)
in \cite{rf:Ikeda}.
As noted in \cite{rf:Ikeda}, the micromagnetism occurs after the formation
of the the heavy fermion state.
Eq.(1) in \cite{rf:Ikeda} should be regarded as the effective Hamiltonian for,
not the bare electrons, but the quasiparticles with the renormalized
interactions, $U$, $V$, $J$.
We can expect that $U$ is renormalized to be the order of the quasiparticle
band-width and $V$,$J<U$ \cite{rf:Torikachvili}.
Then, there exists a stable $d$SDW-region as shown in Fig.1,
even if the possibility of the FM state is included.

In conclusion, the possibility of the FM state modifies the phase diagram
in \cite{rf:Ikeda}.
Since this strong FM enhancement is peculiar to our model,
further careful analyses may be necessary in constructing more realistic
models for URu$_2$Si$_2$.
However, the physical properties of the $d$SDW obtained in \cite{rf:Ikeda}
themselves are not altered at all by the presence of the FM state, so that
the unconventional SDW is still a candidate for the curious magnetism
in URu$_2$Si$_2$.

\vspace{5mm}
\noindent
Hiroaki Ikeda \par
\ \ Department of Physics, \par
\ \ Kyoto University, \par
\ \ Kyoto, 606-8502, Japan \par
\noindent
Yoji Ohashi \par
\ \ Institute of Physics, \par
\ \ University of Tsukuba, \par
\ \ Ibaraki 305, Japan \par

\vspace{3mm}
\noindent
PACS numbers: 71.27.+a, 75.50.Ee


\begin{figure}
\begin{center}
\psbox{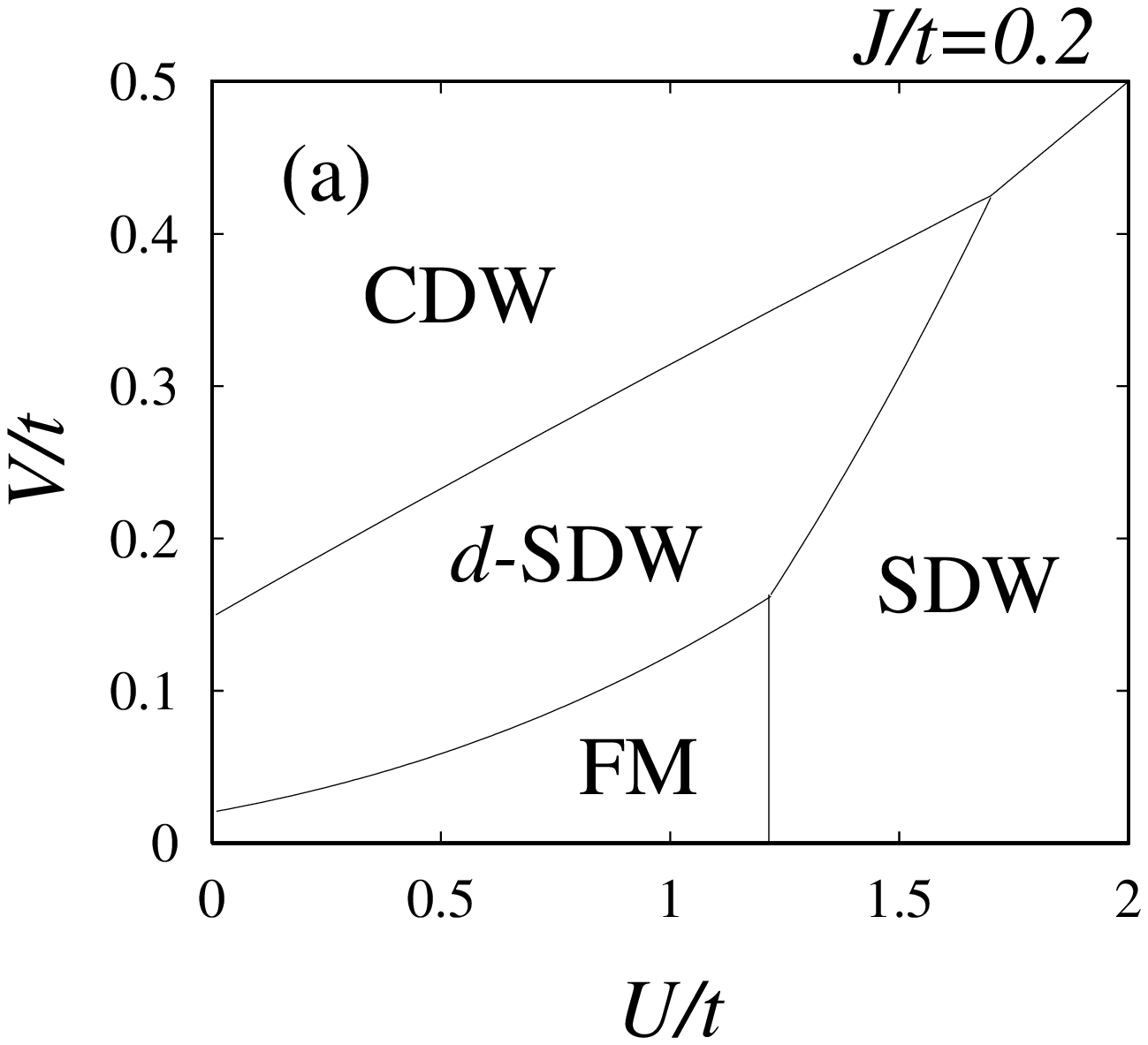,width=150pt}
\psbox{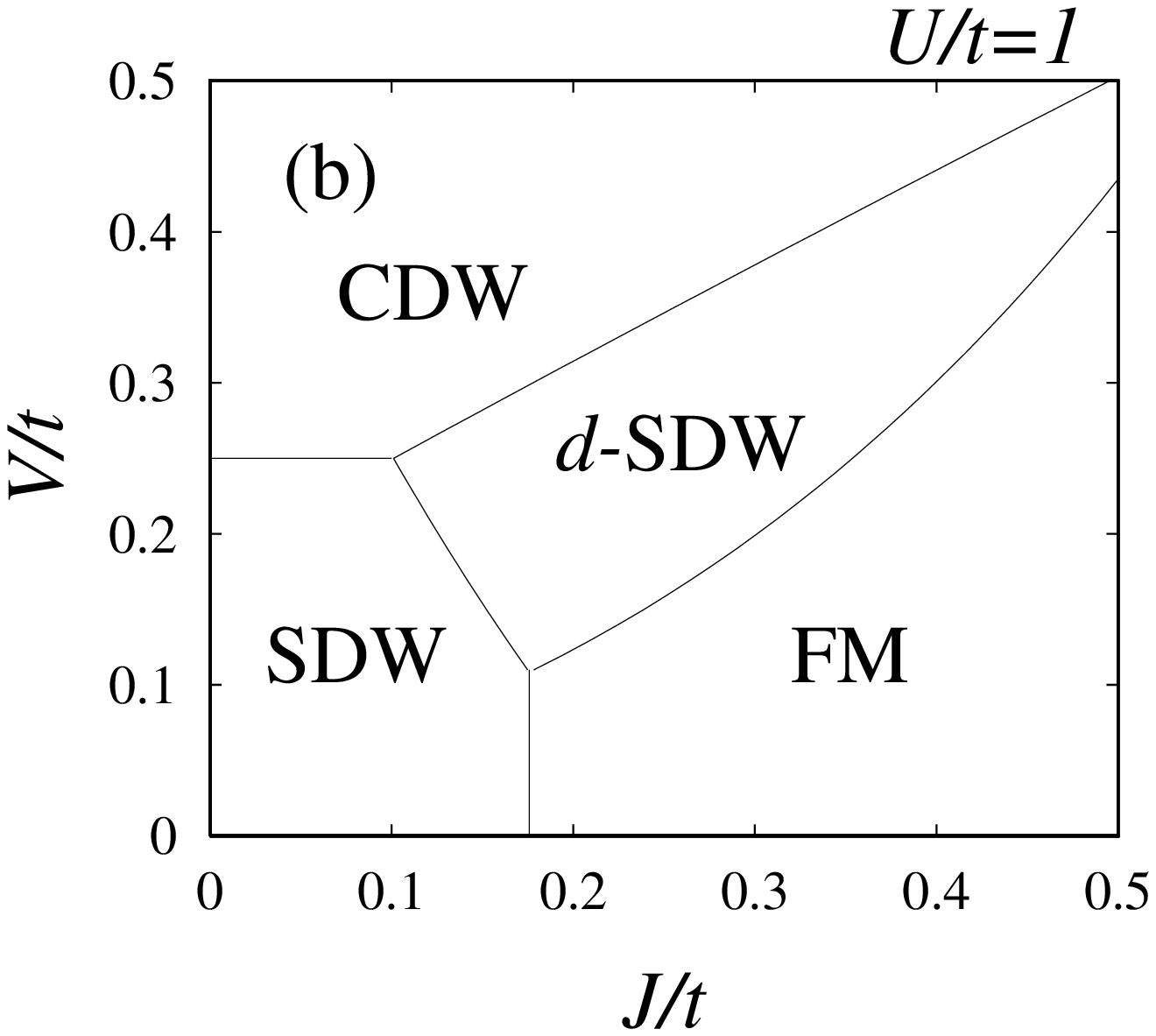,width=150pt}
\end{center}
\caption{(a) $U-V$ phase diagram at $J/t=0.2$. The FM is stable in small $U$
and $V$. (b) $J-V$ phase diagram at $U/t=1$.}
\end{figure}

\end{document}